\documentclass[11pt]{article}
\usepackage[
    top=1in,
    bottom=1in,
    left=0.8in,
    right=0.8in
]{geometry}

\usepackage{graphicx}
\usepackage{xcolor}
\usepackage{titling}
\usepackage{multirow}
\usepackage{float}
\usepackage[skip=5pt]{caption}
\usepackage{enumitem}
\usepackage{url}
\usepackage[T1]{fontenc}
\usepackage{amsfonts}
\usepackage{comment}
\usepackage{subfigure}
\usepackage[
  colorlinks=true,
  linkcolor=blue,
  citecolor=blue,
  urlcolor=blue
]{hyperref}

\usepackage{cite}

\makeatletter
\let\cite@orig\cite
\renewcommand{\cite}[1]{\textsuperscript{\cite@orig{#1}}}
\renewcommand\@biblabel[1]{#1.}
\makeatother

\usepackage{xcolor}
\usepackage{tabularray}
\usepackage{microtype}
\usepackage{ragged2e}
\usepackage{helvet}

\usepackage[T1]{fontenc}
 \newcommand{\tblsize}{\fontsize{6.8pt}{8.4pt}\selectfont}

\usepackage{booktabs} 
\usepackage{array}
\usepackage{longtable}
\usepackage{booktabs}

\title{LLMs in Digital EDA: A perspective on \\shifting roles from Generation to Orchestration}

\author{
    Matthew Youngman,
    Cristian Sestito,
    Themis Prodromakis
}

\date{
    \small
    Centre for Electronics Frontiers, Institute for Integrated Micro and Nano Systems, \\ School of Engineering, The University of Edinburgh, UK \\
}

\begin{document}

\maketitle

\begin{abstract}
Electronic design automation (EDA) has advanced engineering productivity through successive generations of tooling that progressively automate synthesis, optimisation, and verification. Large language models (LLMs) extend this trajectory by enabling direct translation from design intent to hardware implementations. In most of the EDA literature, LLM-based solutions are typically assisting siloed design stages or tasks, however this obscured the drivers by which capability emerges and systems scale. In this Perspective, we instead define three hierarchical roles that reveal how capability accumulates: a Generator that produces design artifacts in a single pass, an Agent that refines outputs through iterative tool feedback, and an Orchestrator that coordinates decisions across EDA-stages. Across published systems, this reveals a \textit{syntax trap} in which models are trained to produce plausible code rather than physically correct hardware, compounded by fragmented tools and loss of design context that obscure how decisions affect later stages. Comparisons across the three roles show that current approaches struggle to scale to industrial designs, motivating a shift towards a standardised, physics-aware orchestrator that connects tools and agents across the EDA flow for more reliable and accessible hardware design.
\end{abstract}

\section{Introduction}\label{Section 1}

The semiconductor industry is facing a productivity gap driven both by global silicon demand, with the semiconductor market projected to exceed \$1 trillion by 2030 \cite{semiconductor_market}, and chip complexity growth. Products now integrate billions of transistors with more diverse architectures and tighter physical constraints \cite{chip_complexity}, all within increasingly short development windows \cite{RevolutionOrHype}. Electronic Design Automation (EDA) tools were introduced to help engineers construct and refine designs using deterministic, physics-based methods \cite{ML4EDA}, improving how design execution is carried out at scale. However, the preceding task of translating engineering intent into formal hardware artifacts remains manual, iterative, and reliant on specialized domain experts who are sparse, concentrated, and difficult to scale \cite{ai_limits_eda, chipchat}, making the production of correct and optimised implementations a key productivity bottleneck.

Large Language Models (LLMs) offer a promising solution to reduce this productivity gap by translating natural language (NL) design intent into structured hardware artifacts through text generation. Digital designs are expressed largely through structured text - hardware description language (HDL) code, specifications, scripts, and constraint files - positioning LLM integration as a natural extension of this \cite{LLMsforEDA, AgenticEDA}. The productivity gains already demonstrated in software engineering, where LLM-assisted development has reduced manual effort and shortened iteration cycles \cite{AIinSWE}, suggest a similar opportunity in hardware, however one constrained by expensive and physics-driven validation. Consequently, two architectural roles have emerged: the \textbf{Generator} produces artifacts from NL in a single forward pass, leaving the engineer to interpret tool output and steer subsequent design iteration manually; the \textbf{Agent} closes this loop by coupling generation to automated tool feedback within a fixed pipeline, enabling automated design repair and optimisation at each cycle. Both roles have produced measurable gains in targeted tasks such as HDL \cite{Towards_Optimal_Circuit_Gen, rtlagent, verigen} and EDA script generation \cite{ChatEDA, ChipNeMo, edaid}, code repair \cite{RTLFixer, craftrtl, aivril2}, verification \cite{veriassist, VeriPrefer}, and hardware dataset construction \cite{Data_is_all_you_need, codegen, deeprtl}. However, important limitations remain, including the difficulty of ensuring physically realisable designs, fragmented reasoning across isolated EDA stages, and increasingly opaque decision-making as systems assume greater autonomy and design scale.

In this perspective, we argue that a third role - the physically-aware \textbf{Orchestrator} - is the next logical step in the progression of LLM roles and the convergent response to the limitations above. Where a generator produces and an agent iteratively repairs, an orchestrator holds control authority over the pipeline itself: it reasons adaptively about which tools to invoke, in what order, and when to redirect effort based on a persistent, cross-stage model of design state, with end-to-end decision provenance emerging as a property of its coordination function. This framing positions the field's capability gains as a consequence of architectural progression - how LLMs are connected to tools and design context, not model improvements alone - and identifies standardised orchestration as the required shift to address hardware's \textit{syntax trap}, EDA stage fragmentation, context statelessness, and explainability defects that current agentic systems cannot inherently resolve.

\section{The Misalignment of LLMs in the Digital Design Flow}\label{Section 2}

\begin{figure}[H]
    \centering
    \includegraphics[width=\linewidth]{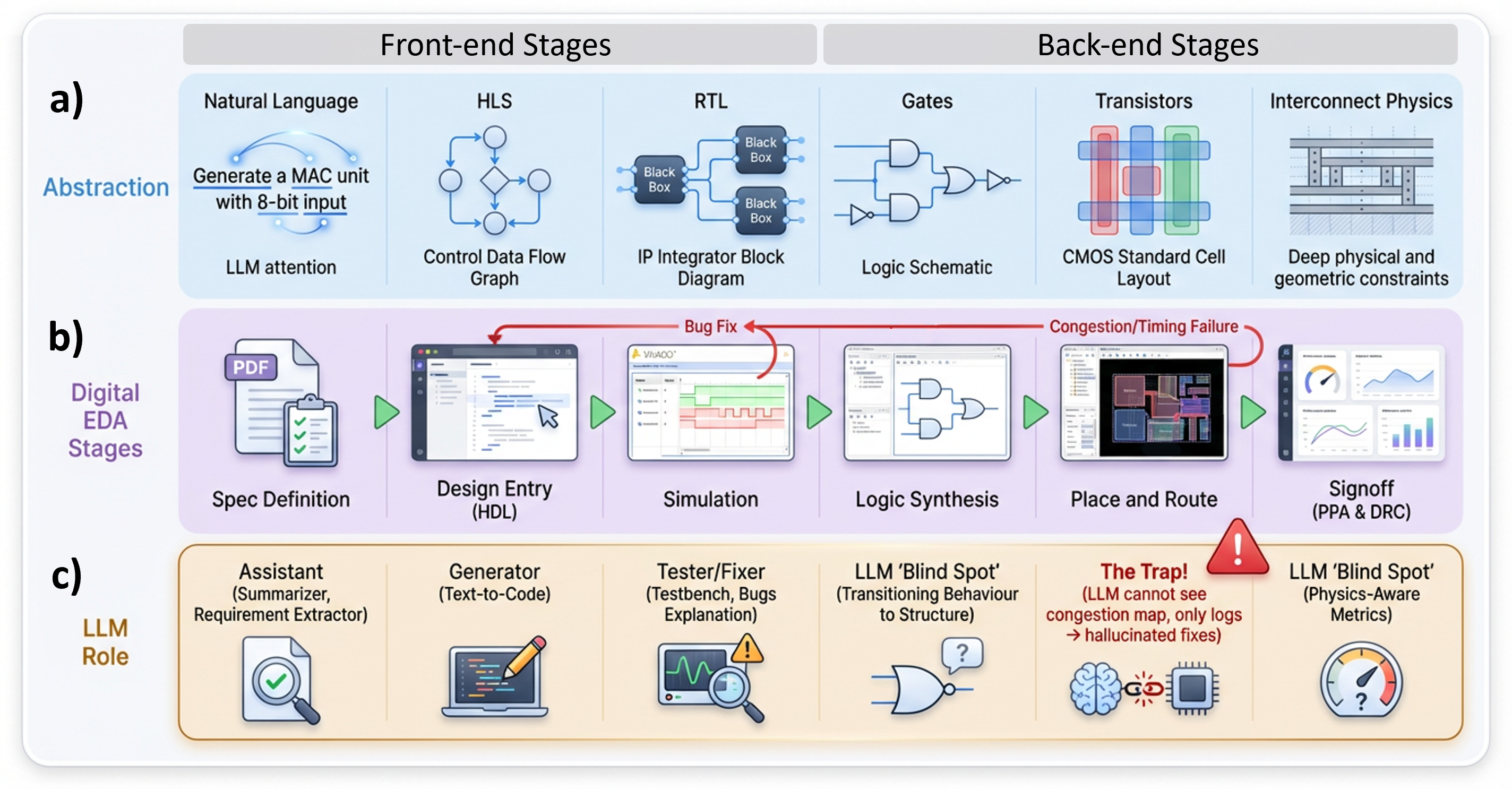}
    \caption{\textbf{Overview of LLMs’ current integration within the digital EDA design flow.} Shows \textbf{a,} technology abstraction levels, \textbf{b,} the sequential digital design stages, and \textbf{c,} the roles LLMs are taking at each step. Identifies “syntax traps” where a lack of physics awareness leads to less effective LLM application.}
    \label{fig:Figure1}
\end{figure}

The digital EDA design flow, illustrated in Figure~\ref{fig:Figure1}, is a staged translation pipeline that converts engineering intent into physics-validated silicon through a series of abstraction levels (Figure~\ref{fig:Figure1}a). Each successive layer — from NL and high-level synthesis (HLS) specifications through register-transfer level (RTL) hardware description to gate-level netlists, transistor standard cells, and physical interconnect layout — allows engineers to express design intent at increasing speed and scale by absorbing the implementation complexity of the level below into automated tooling, producing step-changes in productivity at each transition \cite{ML4EDA, MLforFPGA, ai_supercycle, ai_native_eda}, at the cost of reduced visibility and control over lower-level physical behaviour. Figure~\ref{fig:Figure1}b maps this abstraction hierarchy onto the standard EDA flow, spanning specification definition, design entry, functional simulation, logic synthesis, place-and-route (P\&R), and final signoff. The front-end stages are predominantly language-mediated, where intent is expressed through structured text such as design specifications, HDL code, and constraint files, making them well suited to LLM-based assistance \cite{LLMsforEDA, AgenticEDA}, with gains showcased across NL to specification \cite{ChipGPT} or RTL generation \cite{verigen, Towards_Optimal_Circuit_Gen, GPT4AIGChip}, testbench authoring \& assertion synthesis \cite{veriassist, VFocus, CVDP}, and bug summarisation \cite{rootcause, ChipNeMo, SurveyLLMsEDA}. In contrast, back-end stages transition into physics-enforced representations and deterministic algorithms — timing closure, routing feasibility, and design-rule compliance — with no tolerance for approximation \cite{RevolutionOrHype, AgenticEDA}. LLMs are "blind" to complex physical representations as geometry defies sequential encoding \cite{resbench}, confining back-end LLM contributions to tasks retaining textual character, such as synthesis and P\&R tool-script generation \cite{autoeda, edaid}, EDA log and timing-report interpretation \cite{veriopt, rootcause}, and power, performance and area (PPA) estimation from RTL code \cite{RocketPPA}, and forming a boundary that reflects a more fundamental training-objective misalignment.

LLMs learn through next-token prediction — an objective that rewards linguistic plausibility without any mechanism for training physical correctness — producing an inherent failure mode and misalignment we term the \textit{syntax trap}. This allows LLM generated designs to appear valid under shallow, front-end evaluation yet fail rigorous physical checks at deeper validation levels. For example, a Verilog module may compile successfully and pass simulation while containing subtle timing errors, such as incorrect blocking or non-blocking assignments, that remain invisible to text-level evaluation but can cause functional failure in fabricated hardware \cite{CVDP}. Because these errors can propagate undetected through later design stages, they may ultimately result in silicon re-spins costing millions of dollars and multi-month product delays \cite{SurveyLLMsEDA, RevolutionOrHype}. The trap is quantified across three tiers: at the syntax-level, 55\% of LLM generated Verilog are amenable to automated repairs \cite{RTLFixer}, indicating that token-level plausibility does not reliably translate to syntactically valid hardware descriptions; at the functional level, simulation passage is insufficient, with 44.2\% of testbench-passing designs failing formal equivalence checking \cite{RealBench}; and at the physical level, even functionally correct designs can incur 1.11-1.38$\times$ area overhead relative to human-optimised designs \cite{RTLLM}. This misalignment is reinforced by $pass@k$, the dominant early metric that measures design correctness as the fraction of $k$ samples which compile or simulate correctly \cite{codegen, RTLLM, VerilogEval}, thereby optimising for syntactic and functional plausibility while excluding physical correctness.

The syntax trap persists for two fundamental reasons. First, the physics-grounded data needed to learn hardware behaviour — including synthesis results, timing reports, and physical implementation outcomes — is computationally expensive to generate and orders of magnitude scarcer than conventional training corpora \cite{RevolutionOrHype, Data_is_all_you_need}. Even HDL code remains limited, representing less than 0.1\% of public code corpora due to proprietary IP restrictions and application diversity \cite{ai_supercycle, verigen}. Second, the relationship between HDL and physical outcomes is highly complex and often indirect: small code changes can produce large and uncertain effects on PPA, making physical correctness difficult to infer from text alone.

The field's response has been to shift from single-pass generation (the Generator) toward closed-loop agentic refinement (the Agent), where LLM-generated designs are repeatedly checked by EDA tools and revised using the resulting feedback until quality criteria are satisfied \cite{veriassist, veriopt, MAGE}. This produces a \textit{neuro-symbolic loop}, where LLMs generate candidate designs, while EDA tools enforce physical correctness. Complementary efforts introduced circuit-native intermediate representations (IRs), including abstract syntax trees (ASTs) improving structural code analysis \cite{Veriseek, RTLRewriter} and dataflow graphs (DFGs) enabling signal-level physical reasoning \cite{HDLxGraph, verigrag}, for LLMs to gain beyond-text understanding. For example, tracing a failing signal through a design's parsed syntax tree, rather than searching its text, enabled VerilogCoder's debugging agent to raise pass rates by more than 25 percentage points \cite{VerilogCoder}. However, both agentic systems and IR-enhanced models remain largely confined to individual design tasks and stages (Figure~\ref{fig:Figure1}c), with Agent feedback typically local to the task being completed and IRs capturing structure rather than downstream physical consequences. Consequently, decisions made at one abstraction level remain largely disconnected from their effects elsewhere, motivating the need for \textit{cross-layer understanding} - linking decisions to downstream outcomes - across the entire EDA pipeline \cite{LLMsforEDA}, rather than within the isolated design stages targeted by most current LLM-EDA systems.

\section{Current Applications of LLMs in EDA}\label{Section 3}

\begin{figure*}[t]
    \centering
    \includegraphics[width=\textwidth]{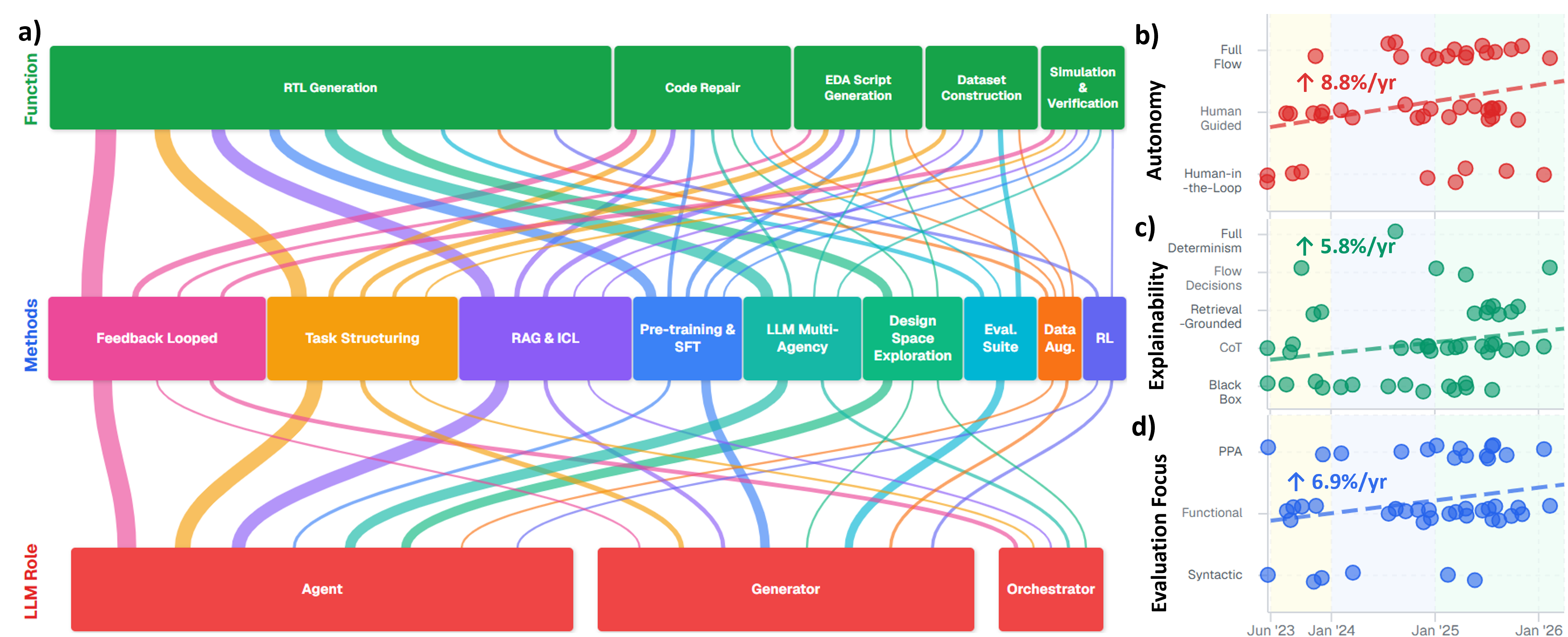}
    \caption{\textbf{Trends in LLM maturity and the landscape of current research. a,} Taxonomy of field across 46 papers, connecting each paper’s EDA function, LLM method stack and LLM role. The progression of LLM capabilities across three axes: \textbf{b,} \textit{autonomy} (degree of agentic workflow control), \textbf{c,} \textit{explainability} (decision provenance), and \textbf{d,} \textit{evaluation focus} (target design challenge ), with annotated trend lines of gradients normalized to axis scale; points sharing the same rank are offset for clarity only. Survey results are in Supplementary Table~\ref{tab:llm_eda_survey} and figure is interactive at \url{https://mattycode101.github.io/LLMs_in_Digital_EDA_Perspective/}.}
    \label{fig:Figure2}
\end{figure*}

To characterise the scope and structure of current stage-local LLM deployments, Figure~\ref{fig:Figure2}a maps the architectures of 46 published LLM-EDA systems across their function, method stack, and architectural LLM role (see Supplementary Table~\ref{tab:llm_eda_survey} and the accompanying online interactive visualisation for the full taxonomy and inter-category relationships). The function distribution highlights the impact of the syntax trap: RTL generation accounts for the largest share of published work, sitting at the design stage where NL and hardware description are closest and where compilation or simulation alone can confirm a plausible result, while back-end physical stages remain sparsely represented. Within the methods, feedback integration, task decomposition, and retrieval-augmented generation (RAG) \& in-context learning (ICL) collectively outweigh pre-training \& supervised fine-tuning (SFT), or reinforcement learning (RL), confirming that current capability gains arise primarily from how LLMs are connected to tools and context rather than how they are trained \cite{AutoChip, AgenticEDA, MCP4EDA}. This trend is reflected in the LLM Role distribution, where Agents now dominate over Generators, marking the field’s shift toward tool-integrated refinement. Orchestrators remain rare, appearing mainly in design space exploration (DSE) and feedback-loop systems, often supported by LLM multi-agency (LLM-MA) frameworks that decompose tasks across multiple general-purpose LLMs \cite{MAGE, edaid, veriopt, aivril2, rtlagent}. However, this decomposition primarily increases parallel execution rather than coordinated control. Roles are typically fixed in advance, interactions follow static handoffs, and no component can adapt the workflow based on global design state \cite{AgenticEDA}, limiting both specialised agent effectiveness and cross-layer understanding.

Figures~\ref{fig:Figure2}b--d characterise the trends of the selected papers along three dimensions using qualitative evaluation rankings to expose structural imbalance in the field's development. Figure~\ref{fig:Figure2}b show Autonomy - how independently a system runs once started - increasing rapidly at an axes-normalised rate of 8.8\% per year. This growth tracks a progression in how models are adapted rather than a uniform rise in raw capability: ICL and RAG at the Generation tier, through SFT on domain-adapted hardware corpora, to RL with EDA tool-derived reward signals at the Agent frontier \cite{ChipSeek-R1, Veriseek, VeriPrefer}. RL represents a notable response to the syntax trap: instead of relying solely on next-token imitation, it optimises models using hierarchical rewards drawn from compilation, functional correctness, synthesis outcomes, and PPA targets, aligning generation with physical outcomes to enable strong performance on curated benchmarks \cite{ChipSeek-R1}. However, because these rewards are defined over fixed reference sets — gold-standard designs \cite{Veriseek}, testbench coverage \cite{VeriPrefer}, or benchmark-specific baselines \cite{ChipSeek-R1} — causing the learning signal to remain tightly distribution-bound. This improves performance within isolated tasks but does not generalise well into cross-stage coordination or deeper links between RTL structure and physical behaviour, leaving the syntax trap largely intact. This rise in autonomy is also not matched by a rise in explainability — how easy it is to trace why a system reached a given decision. Figure~\ref{fig:Figure2}c shows explainability advancing at only 5.8\% per year, a gap that widens as autonomy grows, so that systems make more consequential decisions while the reasoning behind them becomes less accessible. For an industry where formal sign-off requires a verifiable link between each design choice and its physical outcome due to extreme re-spin costs, this divergence is consistently identified as a primary barrier to adoption \cite{AgenticEDA, SurveyLLMsEDA}.

Figure~\ref{fig:Figure2}d reveals a parallel failure for evaluation focus — the design property targeted by a benchmark — which has advanced at 6.9\% per year to remain concentrated on functional correctness, despite industrial competitiveness being primarily determined by PPA. Standard generation benchmarks are correspondingly saturated, with leading systems exceeding 95\% pass@k \cite{verimind, verimind, turtle} on modular tasks within key benchmarks RTLLM \cite{RTLLM} and VerilogEval\cite{VerilogEval}; yet on tasks closer to production workflows — assertion writing, testbench construction, and multi-step agentic design — pass rates fall to roughly 3\% and fail to exceed 34\% \cite{CVDP}. This issue is compounded as designs scale from isolated modules to hierarchical systems. Because LLMs operate within finite context windows, Agent-tier systems must repeatedly reconstruct design context rather than retain it persistently, causing information loss and errors. This lack of persistent memory, termed \textit{statelessness}, disrupts continuity of design state across stages. While manageable for small benchmarks, this becomes increasingly unreliable as module interactions, timing dependencies, and implementation constraints expand the context retention and reasoning requirements, causing performance to degrade sharply on larger designs \cite{AutoSilicon, HiVeGen}, with complete failure reported for designs beyond roughly 3,500 lines of code \cite{RealBench}.

These trends point to fragmentation as a compounding structural limitation: most systems define their own toolchains, feedback loops, and internal representations, preventing cross-system reuse and specialised LLM interaction \cite{autoeda}. The same goal — giving LLMs structural awareness of hardware designs — is pursued through incompatible representations, for example HDLxGraph encodes designs as dual syntax–DFGs for retrieval \cite{HDLxGraph}, while CROP uses dense vector embeddings derived from LLM-generated summaries \cite{CROP}; methods built for one cannot transfer to the other. As a result, partial remedies remain isolated: AutoSilicon and ACE-RTL persist design knowledge across sessions \cite{AutoSilicon, ACE-RTL} to ease statelessness; RTLSquad logs decisions explicitly \cite{RTLSquad} and ChipSeek-R1 exposes reasoning steps \cite{ChipSeek-R1} to ease explainability; and MCP4EDA's typed tool interface improves cross-stage coordination \cite{MCP4EDA}. Yet each fix is system-specific and not reused elsewhere, causing fragmentation to prevent solutions from transferring across systems. Closing these gaps therefore requires not point fixes but a shared architecture with common interfaces, persistent state, and inherent decision traceability.

\section{The Standardised Orchestrator}\label{Section 4}

\begin{figure}[H]
    \centering
    \includegraphics[width=\linewidth]{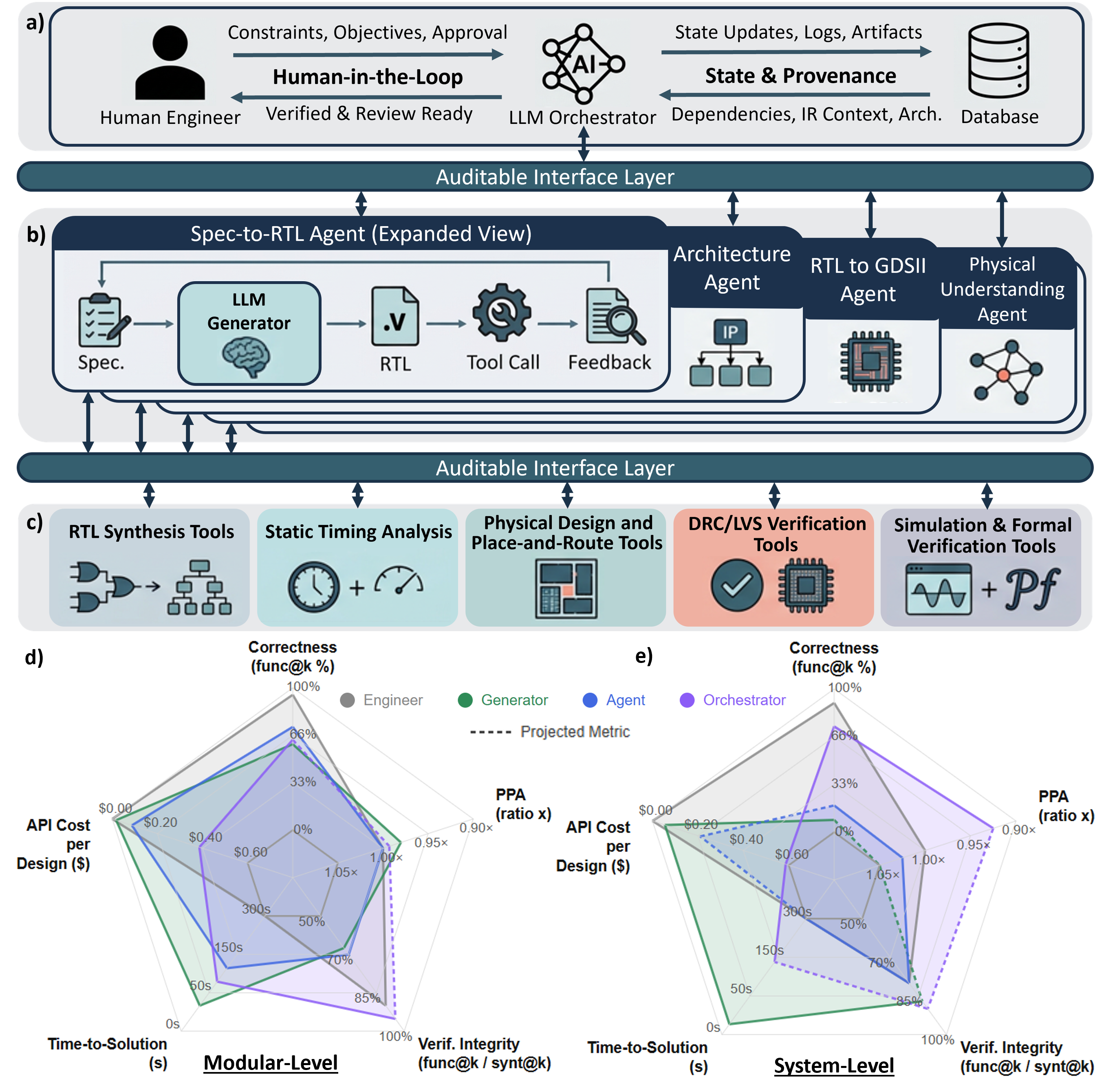}
    \caption{\textbf{Nested LLM architecture framework and multi-axis productivity evaluation}. \textbf{a,} Orchestrator Tier: Collaborative interface between Human, LLM Orchestrator \& Provenance Database. \textbf{b,} Agent Tier: Numerous stateless agents/generators have specifications provided by Orchestrator and execute these in custom pipeline. \textbf{c,} Tool Tier: access to EDA tools for verification of artifacts and design development. \textbf{d, e,} Five-axis productivity comparisons for Spec-to-RTL generation across \textbf{(d)} modular-level ($\leq$150 lines of code) and \textbf{(e)} system-level ($>$200 lines of code) design complexities. Generator, Agent, and Orchestrator LLM architectures are evaluated against human Engineer baselines; see Supplementary Table \ref{tab:radar_data} for plotted results, methodology, and sources.}
    \label{fig:Figure3}
\end{figure}

The architecture that resolves the explainability, statelessness, and fragmentation limitations is best understood not as a new component but as a reorganisation of control. As shown in Figure~\ref{fig:Figure3}a--c, the Generator, Agent, and Orchestrator are nested scopes of reasoning — from individual tasks, to workflows, to EDA multi-stage level coordination. The Orchestrator differs from the Agent through control authority: the capacity to decide which step to take next and deploy numerous agents to complete the decomposed tasks. The limitations identified in the previous section arise largely from the absence of this coordinating role. Once introduced, standardised interfaces reduce fragmentation, persistent design state mitigates context loss, and records decisions and tool interactions to improve provenance. These capabilities are not independent additions, but natural consequences of coordinating the design flow through an auditable layer, enabling LLM-assisted EDA to scale beyond isolated modules toward complete designs.

Because tiers' capabilities are interlinked — tool feedback can improve a strong Generator but cannot compensate for a weak one \cite{AutoChip, rootcause} — orchestration becomes valuable only once Agent-tier systems reach sufficient maturity, helping explain why this transition is emerging now. This moves the research question from generation quality to coordination, with standardisation being the prerequisite for that move. Open protocols, primarily the model context protocol (MCP) and agent-to-agent (A2A) frameworks \cite{Open_Protocols}, replace bespoke tool wrappers and natural-language log parsing with typed, protocol-agnostic interfaces. The effect of this is twofold: improvements compose across systems instead of remaining trapped within them \cite{autoeda}, and provenance becomes a byproduct of structured tool use rather than a post-hoc reconstruction. Applied across the flow, such a layer already returns 15–30\% timing and 10–20\% area improvements by closing the back-end feedback loop and enabling cross-layer understanding \cite{MCP4EDA}.

Control authority also provides an architectural response to the syntax trap and evaluation focus by extending neuro-symbolic feedback beyond isolated optimisation loops. Deterministic EDA tools remain responsible for enforcing correctness, but orchestration ensures that their outputs persist as design context and continuously influence subsequent decisions across the EDA stages. Physical consequences discovered during synthesis or implementation can therefore shape upstream specification or RTL generation rather than appearing only at fixed checkpoints, and can gate each stage's progression with the necessary verification. To reduce verification cost and speed up DSE, large circuit models (LCMs) \cite{ai_native_eda, RevolutionOrHype} may be introduced to act as fast physical surrogates to eliminate poor candidates before tool use, with reported estimators achieving below 2\% power prediction error at more than a 6$\times$ speedup \cite{DeepSeq2}.

Figure~\ref{fig:Figure3}d--e quantifies the performance of each LLM role focusing on the specification-to-RTL generation task across modular and system scales, selected to demonstrate scalability issues as prior work reports sharp degradation in coherence beyond approximately 100--150 lines of generated hardware code \cite{AutoSilicon, RealBench, HiVeGen}. Because hardware-design performance depends strongly on the underlying model, prompting strategy, benchmark, and evaluation methodology, values are aggregated per role across five dimensions: front-end correctness, back-end PPA quality, verification integrity, time-to-solution, and cost per design, with calculation details provided in Supplementary Table \ref{tab:radar_data}. At modular scale, Generator and Agent systems approach human-level correctness while achieving substantially higher productivity in time-to-solution. However, both retain the characteristic signature of the syntax trap: high apparent functional success with weaker verification integrity. The Orchestrator largely closes this gap, albeit at approximately six times the cost, a trade-off that becomes increasingly justified as design risk increases. At system scale, productivity divergence becomes more pronounced: Generator and Agent workflows degrade sharply below 20\% correctness, while Orchestrator performance remains comparatively robust and begins to exceed human baselines in PPA efficiency. The resulting gains are therefore architectural rather than generational, suggesting distinct roles for each paradigm: Generators accelerate design authorship where failure is cheap, Agents support time-consuming iterative refinement, and Orchestrators coordinate system-scale development through cross-stage consistency and validated closure.

\newpage
\section{Outlook on the Industrialized LLM}\label{Section 5}

\begin{figure}[H]
    \centering
    \includegraphics[width=\linewidth]{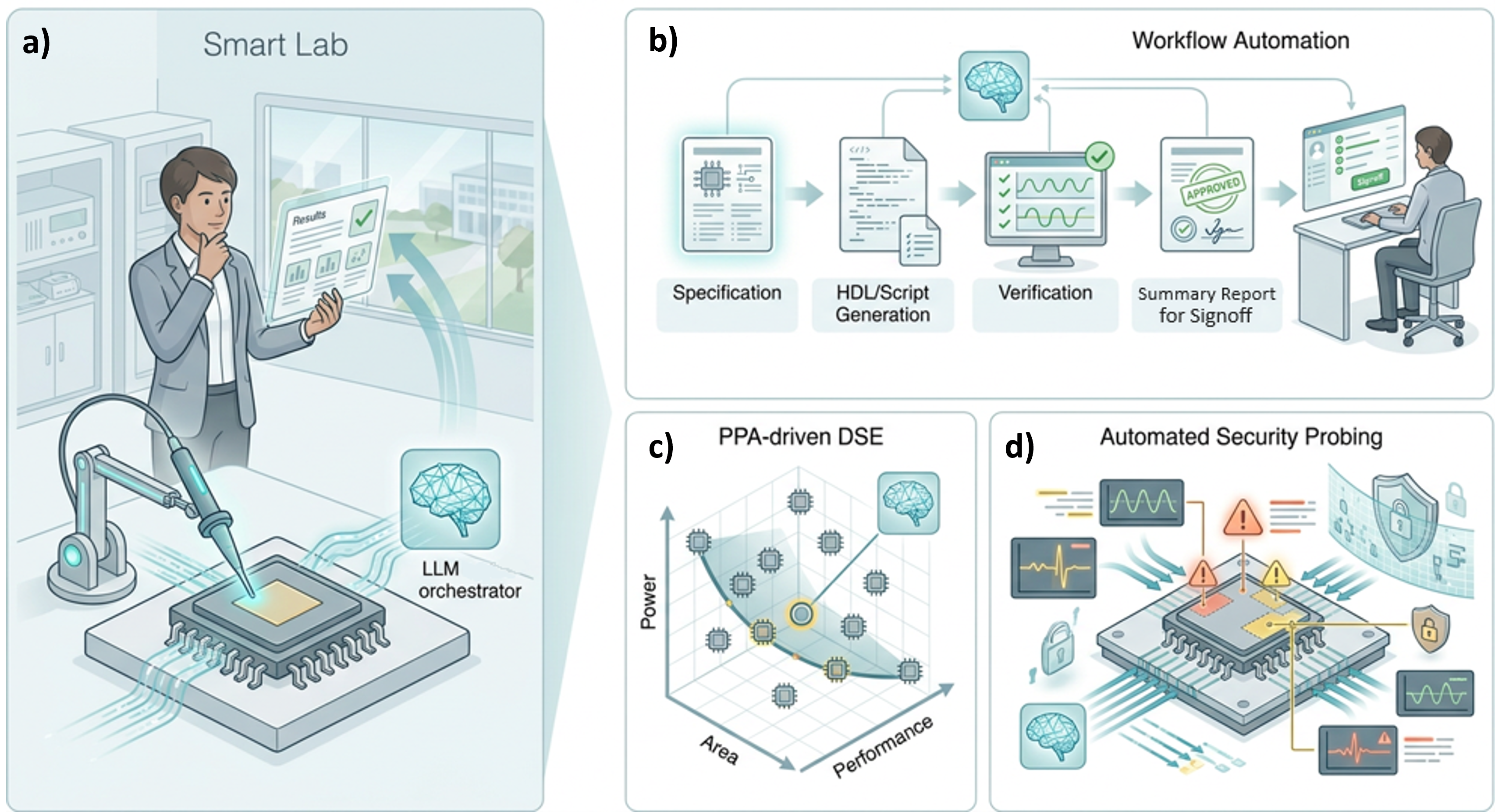}
    \caption{\textbf{Digital design scenarios of Orchestrator} System capabilities radiate from \textbf{a)} The Smart Lab, a central environment for the LLM orchestrator to coordinate physical or simulated experimentation. This environment enables: \textbf{b)} Workflow Automation, support across the EDA flow in near future, shifting engineers towards intent setting and review \textbf{c)} PPA-driven DSE, utilizing agentic search for multi-objective optimization; and \textbf{d)} Automated Security Probing, integrating hardware defensive checks with awareness of LLM adversarial misuse.}
    \label{fig:Figure4}
\end{figure}

The shift towards standardised orchestration this perspective traces is beginning to move from projection into practice, with dominant EDA vendors independently converging on this LLM role. Cadence's ChipStack \cite{cadence_chipstack} delegates across specialised workers behind a persistent IR design model, Synopsys's AgentEngineer \cite{synopsys_agentengineer} wraps autonomous optimisers in a reason–plan–execute–orchestrate loop, and Siemens' Aprisa \cite{siemens_aprisa_ai} embeds auditable, in-flow coordination — the same signature of persistent state, typed coordination, and verification gating the academic field has converged on. Figure~\ref{fig:Figure4} projects where scaled deployment leads, organised around the Smart Lab (Figure~\ref{fig:Figure4}a): a closed-loop environment in which orchestrated agents direct experimentation, receive measured feedback, and return verified candidates to an engineer for oversight. Early forms already exist in closed-loop back-end optimisation \cite{MCP4EDA, RTLRewriter} and in cross-domain refinement spanning digital, analogue, and RF \cite{LaMDA}, with extension to physical prototyping via reconfigurable logic as the next step that would begin to close the Sim-to-Silicon gap \cite{AgenticEDA}. 

Nearest to deployment is workflow automation (Figure~\ref{fig:Figure4}b), which concentrates where correctness signals are cheap and failure is contained. Verification dominates this frontier, accounting for roughly 47\% of engineering effort \cite{wilson_icasic_2024}, with closed-loop systems now achieving 87--90\% Universal Verification Methodology (UVM) coverage \cite{UVM2} and industrial deployments reporting order-of-magnitude reductions in engineering effort across RTL generation, simulation, and formal verification, such as generation of 500+ lined RTL down from 1 week to 1 hour with 28\% lower area \cite{ai_supercycle}. As these tasks become increasingly automated, the engineer's role shifts from authoring code towards specifying intent, defining objectives, and overseeing signoff. Rather than reducing demand for expertise, these gains increase the scope of what engineering teams can practically deliver to match the exponential growth in semiconductor demand.

Enabled by this automation is PPA-driven design-space exploration (DSE) (Figure~\ref{fig:Figure4}c), where the orchestrator's capacity to run many specialised workers in parallel becomes decisive: diversified agents can pursue competing hypotheses across the PPA surface simultaneously. This compresses optimisation cycles that previously demanded extensive manual synthesis effort \cite{Make_Every_Move_Count, Towards_Optimal_Circuit_Gen}. As a result, proficiency in HDL development and EDA scripting becomes less central to productive hardware development, lowering barriers for software-centric researchers and small design teams, and pushing hardware acceleration up the stack. However, this accessibility remains bounded as advanced fabrication still depends on proprietary Process Development Kits (PDKs), large scale design data held by incumbents, and fabrication infrastructure. The opportunity for open ecosystems therefore lies in replicating incumbent resources, and enabling interoperability, portability, and rapid innovation above the fabrication layer.

Greater autonomy is also unavoidably dual-use (Figure~\ref{fig:Figure4}d). The generative capacity that accelerates design also widens the attack surface — Trojan insertion through prompt injection, intellectual-property exposure through cloud inference, and silent functional or side-channel flaws that propagate without formal checking — with reported attack success exceeding 90\% against defensive detection of only 25–35\% \cite{Hardware_Design_Security}. Yet the same orchestration that introduces the risk has the potential to supply the remedy: directed defensively, LLM-based information-flow tracking and systematic adversarial fuzzing expose Trojan sites and illegal states before deployment, while the provenance trace inherent to a standardised, gated flow gives structural resistance to hidden modification \cite{AgenticEDA}. As chiplets and 3D integration multiply design interactions, richer IRs and emerging world models may improve physical reasoning within individual agents, but dependable silicon development increasingly depends on coordinating such capabilities across the design flow.

\section*{Acknowledgements}

This work was supported by the Engineering and Physical Sciences Research Council (EPSRC) AI Hub for Productive Research and Innovation in eLectronics (APRIL) under Grant No. EP/Y029763/1, and by the Royal Academy of Engineering (RAEng) Chair in Emerging Technologies under Grant No. CiET1819/2/93. Figures \ref{fig:Figure1} and \ref{fig:Figure4} were created using Google's Nano Banana Pro and subsequently refined by the authors.

\section*{Competing interests}
The authors declare no conflict of interest.

\newpage
\bibliography{PaperBibtex}

\clearpage

\setcounter{page}{1}

\title{LLM's in Digital EDA: A perspective on \\shifting roles from Generation to Orchestration\\\textbf{Supplementary Information}}

\author{
    Matthew Youngman,
    Cristian Sestito,
    Themis Prodromakis
}

\date{
    \small
    Centre for Electronics Frontiers, Institute for Integrated Micro and Nano Systems, \\ School of Engineering, The University of Edinburgh, UK \\
}

\maketitle

\newpage

\tblsize
\SetTblrInner{rowsep=1.5pt, colsep=3.5pt}

\begin{longtable}{@{}p{1.7cm}p{0.6cm}p{0.95cm}p{2.0cm}p{2.85cm}p{1.23cm}p{1.42cm}p{2.09cm}p{1.42cm}@{}}
\caption{Taxonomy of LLM-for-EDA papers (2023--2026), The table records a manual coding of each system against a fixed schema of categorical (function, method, role) and ordinal (autonomy, explainability, evaluation focus) dimensions, each dimension having a closed set of defined levels; every paper was assigned its single best-fitting level, ordinal dimensions ranked 1…N by increasing capability, and dated by first public release. The growth rates were then obtained by least-squares regression of ordinal level against date, normalised to the axis span (N) and reported as percent of full scale per year (autonomy 8.8\%, explainability 5.8\%, evaluation focus 6.9\%).
\textit{Task abbreviations}: RTL Gen.\,=\,RTL Generation; EDA Scripting\,=\,EDA Script Generation;
Sim.\,\&\,Verif.\,=\,Simulation \& Verification; Dataset Constr.\,=\,Dataset Construction.
\textit{Method abbreviations}: RAG\,\&\,ICL\,=\,Retrieval Augmented Generation / In-Context Learning;
Pre-training\,\&\,SFT\,=\,(Continued) Pre-training \& Supervised Fine-Tuning;
Task Struc.\,=\,Task Structuring; Feedback Looped\,=\,Iterative Feedback Loop;
LLM-MA\,=\,LLM Multi-Agent System; DSE\,=\,Design Space Exploration / Tree Search; RL\,=\,Reinforcement Learning; Data Aug.\,=\,Data Augmentation;
Eval.\,Suite\,=\,Benchmarking / Evaluation Suite.
\textit{Column headers}: \textbf{Ref.}\,=\,Reference; \textbf{Eval.}\,=\,Evaluation Focus.}
\label{tab:llm_eda_survey} \\
\toprule
\textbf{Nickname} & \textbf{Ref.} & \textbf{Date} & \textbf{Design Function} & \textbf{Method Stack} & \textbf{Role} & \textbf{Autonomy} & \textbf{Explainability} & \textbf{Eval.} \\
\midrule
\endfirsthead
 
\multicolumn{9}{l}{\small\textit{(Table~\ref{tab:llm_eda_survey} continued)}} \\[2pt]
\toprule
\textbf{Nickname} & \textbf{Ref.} & \textbf{Date} & \textbf{Design Function} & \textbf{Method Stack} & \textbf{Role} & \textbf{Autonomy} & \textbf{Explainability} & \textbf{Eval.} \\
\midrule
\endhead
 
\midrule
\multicolumn{9}{r}{\small\textit{Continued on next page}} \\
\endfoot
 
\bottomrule
\endlastfoot

Chip-Chat & \cite{chipchat} & May 2023 & RTL Gen. & Task Struc. & Generator & Human-in-the-loop & Chain-of-Thought (CoT) & Syntactic Correctness \\
ChipGPT & \cite{ChipGPT} & May 2023 & RTL Gen. & RAG \& ICL; Task Struc.; Feedback Looped; DSE & Agent & Human-in-the-loop & Black Box Generation & PPA Metrics \\
VeriGen & \cite{verigen} & Jul 2023 & RTL Gen. & Pre-training \& SFT & Generator & Human-guided & Black Box Generation & Functional Correctness \\
RTLLM v1 & \cite{RTLLM} & Aug 2023 & Dataset Constr. & Eval. Suite; Task Struc. & Generator & Human-guided & Chain-of-Thought (CoT) & Functional Correctness \\
ChatEDA & \cite{ChatEDA} & Aug 2023 & RTL Gen.; EDA Scripting & Task Struc.; Pre-training \& SFT; Data Aug. & Agent & Human-in-the-loop & Chain-of-Thought (CoT) & Functional Correctness \\
GPT4AIGChip & \cite{GPT4AIGChip} & Sep 2023 & RTL Gen. & RAG \& ICL; Task Struc.; Feedback Looped & Agent & Human-in-the-loop & Pipeline Stage Collection & Functional Correctness \\
ChipNeMo & \cite{ChipNeMo} & Oct 2023 & RTL Gen.; EDA Scripting; Code Repair & Pre-training \& SFT; RAG \& ICL & Generator & Human-guided & Retrieval-Grounded & Syntactic Correctness \\
AutoChip & \cite{AutoChip} & Nov 2023 & RTL Gen. & Feedback Looped; RAG \& ICL & Agent & Full Flow & Black Box Generation & Functional Correctness \\
RTLFixer & \cite{RTLFixer} & Nov 2023 & Code Repair & RAG \& ICL; Task Struc.; Feedback Looped & Agent & Human-guided & Retrieval-Grounded & Syntactic Correctness \\
VeriPPA & \cite{verippa} & Dec 2023 & RTL Gen.; Code Repair & RAG \& ICL; Feedback Looped; Task Struc. & Agent & Human-guided & Black Box Generation & PPA Metrics \\
Make Every Move Count & \cite{Make_Every_Move_Count} & Feb 2024 & RTL Gen. & DSE; RL; Task Struc. & Agent & Human-guided & Black Box Generation & PPA Metrics \\
Data is all you need & \cite{Data_is_all_you_need} & Mar 2024 & Dataset Constr.; RTL Gen.; Code Repair; EDA Scripting & Data Aug.; Pre-training \& SFT; Task Struc. & Generator & Human-guided & Black Box Generation & Syntactic Correctness \\
VeriSeek & \cite{Veriseek} & Jul 2024 & RTL Gen. & RL; Pre-training \& SFT & Generator & Full Flow & Black Box Generation & Functional Correctness \\
VerilogCoder & \cite{VerilogCoder} & Aug 2024 & RTL Gen.; Code Repair & LLM-MA; Task Struc.; Feedback Looped & Agent & Full Flow & Chain-of-Thought (CoT); Deterministic Trace & Functional Correctness \\
RTLRewriter & \cite{RTLRewriter} & Sep 2024 & RTL Gen. & DSE; RAG \& ICL; Task Struc. & Agent & Full Flow & Chain-of-Thought (CoT) & PPA Metrics \\
CraftRTL & \cite{craftrtl} & Sep 2024 & RTL Gen.; Code Repair & Data Aug.; Pre-training \& SFT & Generator & Human-guided & Black Box Generation & Functional Correctness \\
AutoChip v2 & \cite{autochipv2} & Nov 2024 & RTL Gen.; Code Repair & DSE; Feedback Looped & Agent & Human-guided & Chain-of-Thought (CoT) & Functional Correctness \\
AIvril2 & \cite{aivril2} & Nov 2024 & RTL Gen.; Sim. \& Verif. & LLM-MA; Feedback Looped & Agent & Human-guided & Black Box Generation & Functional Correctness \\
HiVeGen & \cite{HiVeGen} & Dec 2024 & RTL Gen. & Task Struc.; LLM-MA; Feedback Looped; DSE & Agent & Human-in-the-loop & Chain-of-Thought (CoT) & PPA Metrics \\
MAGE & \cite{MAGE} & Dec 2024 & RTL Gen.; Sim. \& Verif. & LLM-MA; Feedback Looped & Orchestrator & Full Flow & Chain-of-Thought (CoT) & Functional Correctness \\
RTL Agent & \cite{rtlagent} & Dec 2024 & RTL Gen.; Code Repair & Feedback Looped; LLM-MA & Agent & Human-guided & Chain-of-Thought (CoT) & Functional Correctness \\
RTLSquad & \cite{RTLSquad} & Jan 2025 & RTL Gen. & LLM-MA; Feedback Looped & Agent & Full Flow & Pipeline Stage Collection & PPA Metrics \\
EDAid & \cite{edaid} & Feb 2025 & EDA Scripting & LLM-MA; Task Struc.; Pre-training \& SFT & Agent & Full Flow & Chain-of-Thought (CoT) & Syntactic Correctness \\
DeepRTL & \cite{deeprtl} & Feb 2025 & Dataset Constr.; RTL Gen. & Pre-training \& SFT; Task Struc.; Eval. Suite & Generator & Human-guided & Black Box Generation & Functional Correctness \\
ResBench & \cite{resbench} & Mar 2025 & Dataset Constr. & Eval. Suite; Data Aug. & Generator & Full Flow & Black Box Generation & PPA Metrics \\
VeriMind & \cite{verimind} & Mar 2025 & RTL Gen. & LLM-MA; Task Struc.; Feedback Looped & Agent & Human-in-the-loop & Chain-of-Thought (CoT) & Functional Correctness \\
TuRTLe & \cite{turtle} & Mar 2025 & RTL Gen. & Eval. Suite & Generator & Human-guided & Chain-of-Thought (CoT) & PPA Metrics \\
CodeGen & \cite{codegen} & Apr 2025 & RTL Gen.; Dataset Constr. & Eval. Suite; Pre-training \& SFT & Generator & Human-in-the-loop & Black Box Generation & Functional Correctness \\
CircuitMind & \cite{Towards_Optimal_Circuit_Gen} & Apr 2025 & RTL Gen. & LLM-MA; RAG \& ICL; Feedback Looped & Orchestrator & Full Flow & Pipeline Stage Collection & PPA Metrics \\
VeriPrefer & \cite{VeriPrefer} & Apr 2025 & Sim. \& Verif. & RL; Pre-training \& SFT; Feedback Looped & Generator & Full Flow & Black Box Generation & Functional Correctness \\
HDLxGraph & \cite{HDLxGraph} & May 2025 & Code Repair; Dataset Constr. & RAG \& ICL; Task Struc. & Generator & Human-guided & Retrieval-Grounded & Syntactic Correctness \\
CVDP & \cite{CVDP} & Jun 2025 & Dataset Constr. & Eval. Suite & Generator & Full Flow & Chain-of-Thought (CoT) & Functional Correctness \\
CROP & \cite{CROP} & Jul 2025 & EDA Scripting & RAG \& ICL; Task Struc.; Feedback Looped; DSE & Agent & Full Flow & Retrieval-Grounded & PPA Metrics \\
ChipSeek-R1 & \cite{ChipSeek-R1} & Jul 2025 & RTL Gen. & Pre-training \& SFT; RL; DSE & Generator & Human-guided & Chain-of-Thought (CoT) & PPA Metrics \\
HW Fail RCA & \cite{rootcause} & Jul 2025 & Code Repair & RAG \& ICL; Eval. Suite & Generator & Human-guided & Retrieval-Grounded & Functional Correctness \\
VeriOpt & \cite{veriopt} & Jul 2025 & RTL Gen. & LLM-MA; RAG \& ICL; Task Struc.; DSE & Agent & Human-guided & Chain-of-Thought (CoT) & PPA Metrics \\
RealBench & \cite{RealBench} & Jul 2025 & RTL Gen. & Eval. Suite; Feedback Looped & Generator & Human-guided & Black Box Generation & Functional Correctness \\
MCP4EDA & \cite{MCP4EDA} & Jul 2025 & EDA Scripting & Feedback Looped; RAG \& ICL; DSE & Orchestrator & Human-guided & Retrieval-Grounded & PPA Metrics \\
AutoEDA & \cite{autoeda} & Aug 2025 & EDA Scripting & Task Struc.; RAG \& ICL & Agent & Full Flow & Chain-of-Thought (CoT) & Functional Correctness \\
AiEDA & \cite{aieda} & Aug 2025 & RTL Gen. & LLM-MA; RAG \& ICL; Feedback Looped; DSE & Agent & Human-guided & Retrieval-Grounded & Functional Correctness \\
VeriPPAv2 & \cite{verippav2} & Sep 2025 & RTL Gen. & Feedback Looped; RAG \& ICL; DSE & Agent & Human-in-the-loop & Chain-of-Thought (CoT) & PPA Metrics \\
VeriGRAG & \cite{verigrag} & Sep 2025 & RTL Gen. & RAG \& ICL & Generator & Full Flow & Retrieval-Grounded & Functional Correctness \\
AutoSilicon & \cite{AutoSilicon} & Oct 2025 & RTL Gen. & LLM-MA; Task Struc.; Feedback Looped & Orchestrator & Human-guided & Retrieval-Grounded & Functional Correctness \\
VFocus & \cite{VFocus} & Nov 2025 & RTL Gen.; Sim. \& Verif. & RAG \& ICL; Feedback Looped; Task Struc. & Agent & Full Flow & Chain-of-Thought (CoT) & Functional Correctness \\
LaMDA & \cite{LaMDA} & Jan 2026 & EDA Scripting; RTL Gen. & Feedback Looped; RAG \& ICL & Agent & Human-in-the-loop & Chain-of-Thought (CoT) & PPA Metrics \\
ACE-RTL & \cite{ACE-RTL} & Feb 2026 & RTL Gen.; Code Repair & Feedback Looped; Pre-training \& SFT; LLM-MA & Agent & Full Flow & Pipeline Stage Collection & Functional Correctness \\

\end{longtable}

\newpage

\begin{longtable}{@{}p{1.0cm}p{1.7cm}p{1.3cm}p{0.9cm}p{0.55cm}p{2.0cm}p{2.5cm}p{2.2cm}p{2.0cm}@{}}
\caption{The table covers specification-to-RTL only (the most measurable, best-evidenced stage), split by benchmark module size into component (\textit{$\leq$150 lines of code}) and system (\textit{>200 lines of code}) tiers. Each value is a multi-source central estimate, pooled across cited studies to hold the task fixed and limit conflation of model, benchmark, and metric, with per-cell provenance in the Notes column; correctness is the median reported \textit{func@1} or \textit{pass@1} (when available), PPA the median geometric mean of normalised area-power ratios against human baselines, verification integrity the ratio of functionally-validated (\textit{form@k}) to synactically correct (\textit{synt@k}) designs, time and cost measure the reported end-to-end automated runtime (\textit{seconds}) and cumulative API expenditure (\$) per design respectively. Unreported cells are estimated from tier and trend, and flagged as projected. Human baselines are fixed by definition where used as references (correctness and verification integrity = 100\%, PPA = 1.00×), and human time and cost are derived from an engineer estimated design effort and hourly rate respectively.
}
\label{tab:radar_data} \\
\toprule
\textbf{Chart} & \textbf{Axis} & \textbf{Tier} & \textbf{Val.} & \textbf{Unit} & \textbf{Papers} & \textbf{Model(s)} & \textbf{Benchmark(s)} & \textbf{Notes} \\
\midrule
\endfirsthead
 
\multicolumn{9}{l}{\small\textit{(Table~\ref{tab:radar_data} continued)}} \\[2pt]
\toprule
\textbf{Chart} & \textbf{Axis} & \textbf{Tier} & \textbf{Val.} & \textbf{Unit} & \textbf{Papers} & \textbf{Model(s)} & \textbf{Benchmark(s)} & \textbf{Notes} \\
\midrule
\endhead
 
\midrule
\multicolumn{9}{r}{\small\textit{Continued on next page}} \\
\endfoot
 
\bottomrule
\endlastfoot
 
\multicolumn{9}{l}{\textit{Modular (Component-Level, $\leq$150 LOC)}} \\
\midrule
 
Modular & Correctness & Human      & 95    & \%       & Estimate & --- & --- & Assumed marginal human error rate \\
Modular & Correctness & Generator  & 59.8   & \%       & \cite{VerilogEval}, \cite{AutoChip}, \cite{rtlagent}, \cite{veriassist}, \cite{AutoSilicon}, \cite{RealBench} & GPT-4, GPT-4o & VerilogEval~\cite{VerilogEval}, RTLLM~\cite{RTLLM} & Median of 6 pass@$k$ \\
Modular & Correctness & Agent      & 72.0   & \%       & \cite{veriassist}, \cite{rtlagent}, \cite{RealBench}, \cite{AutoChip} & GPT-4, GPT-4o & VerilogEval~\cite{VerilogEval}, RTLLM~\cite{RTLLM}, RealBench~\cite{RealBench} & Median of 5 pass@$k$ \\
Modular & Correctness & Orchestrator & 62.9 & \%       & \cite{AutoSilicon} & GPT-4-turbo & AutoSilicon~\cite{AutoSilicon} & Single study \\
\midrule
 
Modular & PPA         & Human      & 1.00   & $\times$ & Definitional & --- & --- & Reference denominator \\
Modular & PPA         & Generator  & 0.98  & $\times$ & \cite{veriassist}, \cite{RTLLM}, \cite{turtle} & GPT-4, ChatGPT-4.0 & VerilogEval~\cite{VerilogEval}, RTLLM~\cite{RTLLM} & Median geomean (Area, Power) \\
Modular & PPA         & Agent      & 1.00   & $\times$ & \cite{veriassist} & GPT-4 & VerilogEval~\cite{VerilogEval}, RTLLM~\cite{RTLLM} & Median geomean (Area, Power) \\
Modular & PPA         & Orchestrator & 0.93  & $\times$ & Projected & --- & --- & Mean of other roles \\
\midrule

Modular & Verif.\ Integ. & Human     & 90   & \%       & Estimate & --- & --- & Residual verification escapes \\
Modular & Verif.\ Integ. & Generator & 66.7  & \%       & \cite{RealBench}, \cite{RTLLM}, \cite{craftrtl}, \cite{verigrag}, \cite{aivril2}, \cite{turtle} & GPT-4, GPT-4o, DeepSeek-R1, Claude 3.5 Sonnet & VerilogEval~\cite{VerilogEval}, RTLLM~\cite{RTLLM}, RealBench~\cite{RealBench} & Median of 10 values \\
Modular & Verif.\ Integ. & Agent     & 70.1  & \%       & \cite{veriassist}, \cite{aivril2}, \cite{verippa} & GPT-4, Claude 3.5 Sonnet & VerilogEval~\cite{VerilogEval}, RTLLM~\cite{RTLLM} & Median of 8 values \\
Modular & Verif.\ Integ. & Orchestrator & 95.2 & \%    & \cite{MAGE} & Claude 3.5 Sonnet & VerilogEval~\cite{VerilogEval} & Median of 2 values \\
\midrule

Modular & Time-to-Sol. & Human     & 300      & s        & Estimate & --- & --- & Conservative lower bound \\
Modular & Time-to-Sol. & Generator & 33.3   & s        & \cite{turtle} & QwQ-32B & VerilogEval~\cite{VerilogEval}, RTLLM~\cite{RTLLM} & Single study \\
Modular & Time-to-Sol. & Agent     & 114    & s        & \cite{HiVeGen}, \cite{veriassist}, \cite{rtlagent} & GPT-4, gpt-4o-mini & Multiplexer, VerilogEval~\cite{VerilogEval} & Median of 4 values \\
Modular & Time-to-Sol. & Orchestrator & 79  & s        & \cite{AutoSilicon} & GPT-4-turbo & AutoSilicon~\cite{AutoSilicon} & Single study \\
\midrule
 
Modular & Cost/Design & Human      & 0.000  & \$       & Definitional & --- & --- & No API cost \\
Modular & Cost/Design & Generator  & 0.017  & \$       & \cite{AutoSilicon} & GPT-4-turbo & AutoSilicon~\cite{AutoSilicon} & Single study \\
Modular & Cost/Design & Agent      & 0.087  & \$       & \cite{HiVeGen} & GPT-4 & Multiplexer & Token-derived \\
Modular & Cost/Design & Orchestrator & 0.386 & \$      & \cite{AutoSilicon} & GPT-4-turbo & AutoSilicon~\cite{AutoSilicon} & Single study \\
 
\midrule\midrule
 
\multicolumn{9}{l}{\textit{System (System-Level, $>$200 LOC)}} \\
\midrule
 
System & Correctness & Human      & 90   & \%       & Estimate & --- & --- & Higher system-level error rate \\
System & Correctness & Generator  & 8.5   & \%       & \cite{RealBench} & GPT-4o & RealBench~\cite{RealBench} & Median of 2 func@1 \\
System & Correctness & Agent      & 18.5  & \%       & \cite{RealBench} & GPT-4o & RealBench~\cite{RealBench} & Median of 2 func@1 \\
System & Correctness & Orchestrator & 73.3 & \%      & \cite{AutoSilicon} & GPT-4-turbo & AutoSilicon~\cite{AutoSilicon} & Single study \\
\midrule

System & PPA         & Human      & 1.00  & $\times$ & Definitional & --- & --- & Reference denominator \\
System & PPA         & Generator  & 1.05   & $\times$ & Projected & --- & --- & Mean of other roles \\
System & PPA         & Agent      & 1.025 & $\times$ & \cite{MCP4EDA} & Claude 4 Sonnet & MCP4EDA~\cite{MCP4EDA} & Geomean (Area, Delay) \\
System & PPA         & Orchestrator & 0.925 & $\times$ & \cite{MCP4EDA} & Claude 4 Sonnet & MCP4EDA~\cite{MCP4EDA} & Geomean (Area, Delay) \\
\midrule
 
System & Verif.\ Integ. & Human     & 80  & \%       & Estimate & --- & --- & Complexity limits coverage \\
System & Verif.\ Integ. & Generator & 87.1 & \%       & \cite{RealBench} & GPT-4-Turbo, o1-preview, Llama-3.1, DeepSeek, GPT-4o & RealBench~\cite{RealBench} & Median of 10 values \\
System & Verif.\ Integ. & Agent     & 80.0 & \%       & \cite{RealBench} & GPT-4o & RealBench~\cite{RealBench} & Median of 2 values \\
System & Verif.\ Integ. & Orchestrator & 90 & \%    & Projected & --- & --- & Estimate ability inherent from arch. \\
\midrule
 
System & Time-to-Sol. & Human     & 300    & s       & Estimate & --- & --- & Conservative lower bound \\
System & Time-to-Sol. & Generator & 13.4 & s         & \cite{AutoSilicon} & GPT-4, GPT-4+CoT & AutoSilicon~\cite{AutoSilicon} & Median of 8 values \\
System & Time-to-Sol. & Agent     & 300  & s         & \cite{MCP4EDA} & Claude 4 Sonnet & MCP4EDA~\cite{MCP4EDA} & Median of 2 values \\
System & Time-to-Sol. & Orchestrator & 138 & s       & \cite{AutoSilicon} & GPT-4+AS & AutoSilicon~\cite{AutoSilicon} & Median of 4 values \\
\midrule

System & Cost/Design & Human      & 0.000 & \$       & Definitional & --- & --- & No API cost \\
System & Cost/Design & Generator  & 0.055 & \$       & \cite{AutoSilicon} & GPT-4, GPT-4+CoT & AutoSilicon~\cite{AutoSilicon} & Median of 8 values \\
System & Cost/Design & Agent      & 0.21   & \$      & Projected & --- & --- & Estimate from other roles \\
System & Cost/Design & Orchestrator & 0.587 & \$     & \cite{AutoSilicon} & GPT-4+AS & AutoSilicon~\cite{AutoSilicon} & Median of 4 values \\
 
\end{longtable}

\end{document}